\documentclass[twocolumn,pre]{revtex4}
\usepackage{bm}
\usepackage{amsmath}
\usepackage{amsfonts}
\usepackage[latin1]{inputenc}
\usepackage[10pt]{type1ec}
\usepackage{calligra}
\usepackage[T1]{fontenc}
\usepackage[polutonikogreek,italian,english]{babel}
\newcommand{\dif}{\mathrm{d}}
\newcommand{\blangle}{ \big\langle }
\newcommand{\brangle}{ \big\rangle }
\newcommand{\equal}{\stackrel{\mathrm{def}}{=}}
\newcommand{\barra}{\mathrm{bar}}
\newcommand{\ther}{\mathrm{ther}}
\newcommand{\inter}{\mathrm{inter}}
\newcommand{\syst}{\mathrm{syst}}
\newcommand{\tot}{\mathrm{tot}}
\newcommand{\neqq}{\mathrm{neq}}
\newcommand{\eq}{\mathrm{eq}}

\begin{document}

\title{Statistical Thermodynamics\\ 
for metaequilibrium or  metastable states}

\author{A.~Carati}
\email{andrea.carati@unimi.it}
\affiliation{Department of Mathematics, Universit\`a degli Studi di  Milano - 
                Via Saldini 50, I-20133 Milano, Italy}
\author{A. Maiocchi}
\email{alberto.mario.maiocchi@gmail.com}
\affiliation{Department of Mathematics, Universit\`a degli Studi di  Milano - 
                Via Saldini 50, I-20133 Milano, Italy}
\author{L. Galgani } 
\email{luigi.galgani@unimi.it}
\affiliation{Department of Mathematics, Universit\`a degli Studi di  Milano - 
                Via Saldini 50, I-20133 Milano, Italy}

\date{\today}

\begin{abstract}
We show how statistical thermodynamics can be formulated 
in situations in which thermodynamics applies, while  equilibrium
statistical mechanics does not. A  typical case is, in the words of  
Landau and Lifshitz, that of partial (or incomplete) equilibrium. One has
a system of interest in equilibrium  with the environment, and  
measures  one of its quantities, for example its specific heat, by
raising the temperature of the environment.  
However, within the
observation time the global 
system   settles down to  a state of apparent equilibrium, so that the 
measured value of the specific heat is  different from  the
equilibrium one.
In such cases {formul\ae} for quantities such as the
effective specific heat exist, which are provided by Fluctuation Dissipation
theory. However, what is lacking is a proof that internal energy
exists, i.e.,  that the fundamental
differential form $\delta Q-\delta W$ (difference between the heat absorbed
by the system and the work performed by it) is closed. Here we show
how the coefficients of the fundamental form can be expressed in  such a
way that the closure property of the form becomes manifest, so that the
first principle is proven. We then show
that the second principle too follows, and indeed as a consequence 
of microscopic time--reversibility. The treatment is given in a
classical Hamiltonian setting. One has a global time--independent
 Hamiltonian system constituted by the system of interest and two auxiliary ones
controlling temperature and pressure, and the occurring of a process
due to a change in the thermodynamic
parameters is implemented  by a suitable choice of the measure for
the initial data.
\end{abstract}



\maketitle

\section{Introduction}\label{sec1}
There are situations of
 metaequilibrium or metastability, in which the laws of
phenomenological thermodynamics apply, whereas  equilibrium
statistical thermodynamics 
 does not.  Here we show how at least a first step can be made in
 providing a formulation of statistical thermodynamics for such cases,
working in a classical Hamiltonian setting.

A paradigmatic case  is  that of partial (or incomplete) equilibrium, 
as Landau and Lifshitz  call it (see \onlinecite{landau}, sec. 4). 
Macroscopically one  has a  system  characterized by a 
phenomenological  internal energy, say
$U (T,p)$ if its thermodynamic  state is defined 
by temperature $T$ and pressure $p$.
Microscopically one has a  corresponding dynamical 
system with a set of degrees of
freedom, a Hamiltonian $H$ and a given family of invariant
probability measures, for example the generalized Gibbs measure with
parameters $T$ and $p$.   At equilibrium one makes the identification 
$$
U=\langle H\brangle_{T,p}\ ,
$$ 
where $\langle \cdot \brangle_{T,p}$ denotes  mean with respect 
 to the  invariant generalized Gibbs measure for the given values of
 $T$ and $p$. 
In the   case of partial (or incomplete) equilibrium, however,
up to a given observation time $\tau$ only a fraction of the degrees
of freedom   is actually  able to  react to  changes 
of the thermodynamic parameters. 

Thus, considering for
example the case of the  specific heat at constant volume 
$C_V$,  this   is  a well defined thermodynamic 
quantity which has  a concretely
measured value,  but this value  does  not agree with the one predicted
by equilibrium statistical mechanics. 
A typical case, mentioned  by Landau and Lifshitz,  is \emph{``the partial
equilibrium of a mixture of several substances which interact
chemically. Owing to the relative slowness of  chemical processes,  the
equilibrium connected with the  motion of molecules occurs,  in
general, considerably sooner  than the equilibrium connected with the 
interchange of atoms between the molecules, i.e., connected with the
composition of the mixture''}.
For example,  in a system composed of a mole of graphite $C$ and a mole
of $O_2$, at room temperature combustion does not occur, and the
measured heat
capacity at constant volume is simply the sum of the specific heats of
the two components, i.e., $1R +(5/2) R=(7/2)R$, whereas at equilibrium
the heat capacity is that of one mole of carbon dioxide $CO_2$, namely,
$(5/2) R$. Another situation presenting some analogy with 
partial equilibrium is that of 
the \emph{``time dependent specific heat''} of Birge
and Nagel \cite{birge}.

On the other hand, in all the cases mentioned there exists
a theoretical formula for  $C_V$ which explicitly
depends on the observation time, namely, the formula  provided by
Fluctuation Dissipation (FD) theory, which is thus  a good candidate for a
theoretical counterpart to the  phenomenological value.   
More in general, in situations of  metaequilibrium 
 one has  available  the differential forms $\delta  Q$ and $\delta W$ 
of  the heat absorbed by the system and of the work performed by it, 
with  coefficients which have well defined phenomenological  values. 
Analogously,
the corresponding theoretical coefficients provided by FD theory have
well defined values too,  if a suitable    dynamical
relaxation  did occur, albeit possibly a partial  one.

However,  there remains open the problem  of proving
that a consistent statistical thermodynamics can be formulated.
 At a phenomenological level, the internal energy can be defined
by integration of the available differential form $\delta Q -\delta
W$, which is closed   in virtue of the  first
principle. At a theoretical level,
instead, such a property should be proved, but a proof is  apparently
lacking.  

In this  paper we  show that, in situations of metaequilibrium,  the
formul\ae\ for the coefficients of the fundamental form
$\delta Q-\delta W$ provided by FD theory can be expressed in such a
way that the closure property of the form becomes manifest. This
proves the first principle.
Moreover, the second
principle too is  proven, as a consequence of microscopic
time--reversibility.
 So one would  have a complete
 statistical thermodynamics if one could also prove the validity of
the zeroth principle (transitivity of metaequilibrium), or of a 
suitable generalization of  it. In any case one has available a candidate
for the internal energy, namely,  the state function $U$ which is 
obtained by  integrating the  form $\delta Q-\delta W$, 
with  coefficients  provided by FD theory. 
However, and this is perhaps the more relevant  contribution of this
paper,
it turns out that in general such an  internal energy 
$U$   cannot be expressed  as the mean of the Hamiltonian with
respect to the invariant measure considered. 

The main point thus consists in introducing  a suitable definition for 
the statistical mechanical  analogues of  $Q$ and $W$, 
as quantities parametrically dependent on
the observation time $\tau$ (which we think of as fixed once for all,
and will be mostly left understood). To this end we essentially follow
the procedure used in FD theory, which in turn mimics the procedure 
of phenomenological thermodynamics, inasmuch as  it  takes  into account only 
the fraction of energy which is actually exchanged with the
measurement apparatus,  
when an increment is given to the thermodynamic parameters $T$ and $p$ 
of the latter.  
It is left to the dynamics itself, which involves the
interactions between the system of interest and the measurement apparatus,
to  determine the amount of energy which is actually
exchanged. It will be seen, however, that the definitions given here
are not exactly those  of FD theory. Some criticism was already
raised against the latter \cite{nielsen}. The present  definitions
are in our opinion somehow simpler,  and more suited to the 
aims of thermodynamics.

In order to implement the idea of mimicking  the procedure used in
phenomenological thermodynamics,  we have first of all 
 to consider the system of interest as coupled
with two auxiliary ones, a thermostat which controls the temperature
and a barostat which controls the pressure.
The global system is  studied as an isolated, time independent,   
Hamiltonian system. The existence of different  states
corresponding to different values of the macroscopic  thermodynamic
parameters $T$ and $p$ is taken into account
by introducing different choices for the probability measure of the
initial data  in the global phase space. This way of handling the
problem  will make the
proofs of the first and of the second principles particularly simple.

In section \ref{sec2} the analogue of the fundamental form is
defined and the
first principle is proven. The second principle is proven in
section \ref{sec3}. To this end, making use of microscopic
time--reversibility,  we preliminarily
 establish  {formul\ae} of FD theory type, 
which express  the coefficients of the form $\delta Q$ in terms  of
equilibrium correlations of suitable functions. From this, in
particular, the property $C_p\ge 0$ follows, which amounts to the
classical formulation of the second principle in the form of Clausius. In
section~\ref{sec4}  it is shown how the thermodynamic temperature of the
system of interest  coincides with that of the thermostat if a
suitable dynamical  relaxation takes place.
Further  discussions are given in the conclusive section. 

In an appendix
it is  discussed how the present results may be of interest for the
Fermi--Pasta--Ulam problem, the analogy of which with glassy--like
systems was first proposed by a group of people about Parisi in the
year 1982 \cite{parisi2} (see also \onlinecite{parisi3} and the review 
\onlinecite{fpurev}). 

\section{The model,  Microscopic analogue of the fundamental form, 
and Deduction of the first principle}\label{sec2}
As usual (see \onlinecite{fermi}, or  \onlinecite{khinchin}, page
131), we  take the 
system of interest (say, a fluid)   enclosed in a cylindrical
vessel one basis of which is fixed, while the other one is a movable
piston. The distance of the piston from the fixed wall will be  
denoted by $l$, so that
to any distance $l$ there corresponds a volume $ V=
{ A}l$ of the system of interest, ${ A}$  being the 
area of the piston. Through the two bases of the cylinder the 
system of interest can exchange energy with two other systems, a
thermostat  which
acts as a source of heat (through the fixed basis), and a barostat
which acts  as a source of purely
mechanical, or adiabatic,  work (through the movable piston).  
It is well known \cite{sinai} that in dynamical terms
the adiabaticity condition  is very well satisfied up to extremely
long times  if the piston is modeled as  a single rigid particle of 
macroscopic mass. The two auxiliary systems are thought of  as having 
no direct mutual interaction.

The global  system (system of interest plus sources of heat and of
work) is  dealt with in a classical Hamiltonian setting, with a given
time independent  Hamiltonian
and a fixed global  geometry, the position of the piston being
considered as one of the configurational coordinates of the global system. 
One thus  has a global phase space with
canonical coordinates
$$
z=(x,x',x'', l, p_l)
$$
where $x$ are the canonical coordinates of the  system of interest,
$x'$ those of the heat source, $x''$ those of the work source, $l$ 
the distance of the piston from the fixed wall,  and $p_l$ the 
corresponding conjugate momentum.
The total Hamiltonian is decomposed as 
$$
H_{\tot}(z)= H(x,l)+ H_{\ther}(x')+H_{\barra}(x'',l)+ H_{\inter}(z) \ .
$$
Here $H$ is the partial Hamiltonian of the
system of interest, while $H_{\ther}$, and $H_{\barra}$ are
the partial Hamiltonians of the sources, the labels \emph{ther}
 and \emph{bar} standing for thermostat and barostat respectively.
 Finally there appears  an
 interaction Hamiltonian, which should contain a term of  the form
$p_l^2/2M$ for the kinetic energy of the piston, dealt with as a
 single rigid particle of mass  $M$ of macroscopic size.
As usual it is understood that, for large systems, 
 the  interaction Hamiltonian,
although playing an essential dynamical role in allowing for energy
exchanges, can be neglected in estimating mean values of the
quantities of interest. Indeed 
such quantities are 
proportional to the volume of the system of interest, whereas  the
interaction terms are proportional to the area of the basis of the
cylinder; furthermore,  motions are considered in which the
piston moves very slowly.
The flow in the global phase space, induced by the global Hamiltonian $H_{\tot}$,
will be denoted by $\Phi^t$.

Our aim is to estimate the energy exchanges between the
system of interest and each of the sources (namely, the heat $Q$ absorbed
by the system and the work $W$ performed by it). This involves first
of all the dynamics of the global system, which induces corresponding
exchanges  for any initial datum $z$, and then the specification of
the measure over the initial data which should be assigned in order to estimate
mean values.  

For what concerns the dynamics, along any solution of 
the equations of motion  the
total energy is conserved, so that, neglecting the contribution of the
interaction Hamiltonian, one has energy conservation in the form 
\begin{equation*}
\left[H+H_{\ther}+H_{\barra}\right](\Phi^\tau z)=
\left[H+H_{\ther}+H_{\barra}\right](z) \ ,
\end{equation*}
or equivalently
\begin{equation}\label{consenergia2}
[\Delta H+\Delta H_{\ther}+\Delta H_{\barra}](z)=0\ ,
\end{equation}
where we have introduced the notation
\begin{equation}\label{deltaf}
\Delta F(z)\equiv \Delta^\tau F(z)\equal F(\Phi^\tau z)- F(z)\ 
\end{equation}
for the change of any dynamical variable  $F$  induced, at the
observation time $\tau$,  by the
time evolution $\Phi^t$ of the global system, as a function of the initial
point $z$ of phase space.  Relation (\ref{consenergia2})
will finally lead to a deduction of the first principle for the
energy exchanges.  This, however, requires to previously define the
measure on the global phase space that should be used in estimating 
mean values,  or even  typicality of an orbit.

Clearly, invariant measures correspond to  equilibrium states.
Indeed  the  mean values of dynamical variables not explicitly
depending on time,  computed with respect to an invariant measure, 
turn out to be time independent. Thus the mean values of the three
energies $H$, $H_{\ther}$ and $H_{\barra}$ do not change with time, i.e., 
there are no energy exchanges, notwithstanding the fact that no one of
the three partial Hamiltonians is an integral of motion.
 Choosing a measure which is not
invariant, leads instead to a thermodynamic process. We are interested
in processes induced by an increment  of the thermodynamic
parameters of the sources, with respect to those characterizing 
an invariant measure.  

To this end we consider  a family of measures  
relative to the global
system. The family  depends on two ``external'' thermodynamic
parameters relative to
the sources (say temperature $T$ for the heat source and pressure $p$
for the work source), and furthermore on at  least
two  ``internal''  parameters, say
$\alpha=(\alpha_1,\alpha_2)$, relative to the system of interest. So
the family will have   densities of the form
$$
 \rho(T,p, \alpha; z)\ .
$$

We then assume that, for any pair $T,p$ of external thermodynamic
parameters, there exists an equilibrium   
state of the global system (that might be  called \emph{reference state}), 
which corresponds  to some definite values $\overline 
\alpha=\overline \alpha(T,p)$ of the internal parameters $\alpha$ for
which the measure is invariant. In particular the invariant measure
could be a global Gibbs one. This is the typical case one considers
in discussing partial equilibrium. The more general case considered
above is of interest when discussing metastability. 

Given one  such equilibrium  state with density 
\begin{equation}\label{notazione2bis}
\rho(T,p,\overline \alpha ;z)\ ,
\end{equation}
we then introduce the state in which the external parameters have been 
incremented, \emph{while  the  values of the internal parameters  
are kept constant}. 
In general such a measure will  not be   invariant, i.e., will
correspond to a nonequilibrium situation,  
leading to a process. 
The corresponding measure  will have   a  density  
\begin{equation}\label{notazione2}
 \rho(T+\dif T,p+\dif p,  \overline \alpha; z)\ .
\end{equation}
Notice that, in the present model, 
 a change of the external parameters can be obtained
without any alteration of the global geometry of the system, because a
change in the pressure $p$ of the barostat can be produced by a
suitable change of its temperature, which in turn can be
described through a change of the measure on the initial data.

Mean values with respect to  the
noninvariant measure with density (\ref{notazione2}) will be denoted by 
$\langle \cdot\brangle^{\neqq}$,  while those referring to the
invariant  measure with density (\ref{notazione2bis}) might be denoted by
$\blangle \cdot\brangle^{\eq}$, and instead will be simply denoted by 
$\blangle\cdot\brangle$:
\begin{equation}\label{notazionimedia}
\begin{split}
\blangle F\brangle^{\neqq}&\equal \int F(z)\, \rho(T+\dif T,p+\dif p,  
\overline \alpha; z)\, \dif z\ , \\
\blangle F\brangle&\equal \int F(z)\, \rho(T,p,  
\overline \alpha; z)\, \dif z\ .
\end{split}
\end{equation}
 
Let us now take as a  measure for the initial data of the global
system the noninvariant one with density (\ref{notazione2}).
For any initial datum $z$,
which evolves to $\Phi^\tau z$, there remain defined an increment of
the energy of the thermostat and an increment of the energy of the
barostat, and this  naturally leads to  microscopic  definitions of 
the heat  $Q$ absorbed by the system of interest and of
the work $W$ performed by it. Indeed,  by analogy with the
standard  equilibrium procedure, for any given dynamical variable
$F(z)$  one considers the
increment  corresponding to each initial datum, 
and then takes the mean value over  the data
distributed according to the given  (nonequilibrium) measure. 
As usual, one might presume that
such mean values are good representatives of the values actually taken
in correspondence to single typical orbits.  

So, recalling the meaning  (\ref{deltaf}) of $\Delta F$, we
introduce the  \emph{definitions} which constitute the key
point of the whole paper, namely,
\begin{equation}\label{trenuova}
\begin{split}
 Q&\equal -\,  \blangle \Delta H_{\ther}\brangle^{\neqq} \equiv\\ 
 -\, &\int\left[ H_{\ther}(\Phi^\tau z)-H_{\ther}(z)\right]\,
\rho(T+\dif T,p+\dif p,  \overline \alpha;z)\,   \dif z\  ,\\  
W& \equal\blangle \Delta H_{\barra}\brangle^{\neqq}\ .
\end{split}
\end{equation}

We can now come to the first principle.
From the definitions (\ref{trenuova}) of $Q$ and $W$, and from energy  
conservation 
(\ref{consenergia2}), one gets
\begin{equation}\label{primo}
\begin{split}
 Q-W &= \blangle \Delta H(z)\brangle^{\neqq}\\
& \equiv \int \Delta H(z)\,
\rho(T+\dif T,p+\dif p,  \overline \alpha;z)\,   \dif z\ . 
\end{split}
\end{equation}
Then, by  expanding $\rho$ to first
order in $\dif T$ and $\dif p$, differential forms $\delta Q$ and $\delta
W$ remain thus defined. So the
analogue of the fundamental form $\delta Q-\delta
W$ turns out to be  expressed in terms of the
Hamiltonian  $H$ of the system of interest only (although still in terms of a
measure involving the two sources),  being given by
\begin{equation}\label{formafond}
\delta Q-\delta W= c_1(T,p) \,\dif T +c_2(T,p)\, \dif p\ , 
\end{equation}
with coefficients 
\begin{equation}\label{cinque}
\begin{split}
c_1&\equal \int \Delta H(z)\
\frac{\partial}{\partial T}\rho(T,p,\overline\alpha; z) \dif z\ ,\\
c_2&\equal \int
\Delta H(z)\
\frac{\partial}{\partial p}\rho(T,p,\overline\alpha; z) 
\dif z\ .
\end{split}
\end{equation}

Proving the first principle amounts to proving that the  differential
form $\delta Q-\delta W$ is closed. On the other hand, the variables 
$T$ and $p$ enter the integrals at the right hand side of (\ref{cinque}) 
only through  the  factor $\frac{\partial}{\partial T}\rho$ or
the factor $\frac{\partial}{\partial p}\rho$. Thus, assuming 
that the density $\rho$ is smooth enough in the parameters, one
has 
$$
\frac{\partial c_1}{\partial p}= 
\frac{\partial c_2}{\partial T}\ ,
$$
which is the analogue of a Maxwell relation. So  the form 
$\delta Q-\delta W$ is closed,
and  there exists (at least locally)  a function
$U=U(T, p)$ such that $\delta Q-\delta W=\dif U$. So the internal energy
can  be defined  by
integrating the differential form (\ref{formafond}). The
first principle is thus  proven. 
Notice that the proof
is very simple due to the fact that we are considering a closed model,
in which the changes in the external parameters  involve only changes
in the measures, and not in the dynamics (at variance with what
 occurs in standard FD theory).

We close this section with two  remarks.  The first one
concerns the fact that the internal energy $U$
cannot be expressed as the mean of the
energy $H$ with respect to the invariant measure. For example,
the state function $\tilde U$ defined by 
$$
\tilde U=\int H(z)\, \rho (T,p,\overline \alpha; z)\dif z 
$$
produces a differential form
\begin{equation}\label{formafondfasulla}
\dif \tilde U= {\tilde c}_1(T,p) \,\dif T +{\tilde c}_2(T,p)\, \dif p\ , 
\end{equation}
with coefficients 
\begin{equation}\label{cinquefasulla}
\begin{split}
{\tilde c}_1&\equal \int  H(z)\
\frac{\partial}{\partial T}\rho(T,p,\overline\alpha; z) \dif z\ ,\\
{\tilde c}_2&\equal \int
 H(z)\
\frac{\partial}{\partial p}\rho(T,p,\overline\alpha; z) 
\dif z\ 
\end{split}
\end{equation}
which, at variance with (\ref{cinque}), involve integrals of $H$
rather than of $\Delta H$.

The second remark  concerns the
relation  between 
the dynamical  increment of internal energy 
\begin{equation}\label{energiainterna}
\Delta^\tau U\equal Q-W
\end{equation} 
defined  through (\ref{primo}) -- in correspondence 
to increments $\dif T$,  $\dif p$ of the parameters --  
and the increment of the equilibrium internal energy which
is obtained by incrementing the parameters $T$ and $p$ in the
equilibrium  measure. This relation is obtained  by remarking
that  the definition
(\ref{primo}),  which involves the time evolved $H(\Phi^\tau z)$
of the dynamical variable $H$ 
 while keeping  fixed the measure, can also be dually expressed
by letting the measure evolve with time, while keeping fixed the  dynamical
variable  $H$:
\begin{equation}\label{primoaltroaltro}
\begin{split}
 \Delta^\tau U = \int  H(z)\,
\Big[ &\rho(T+\dif T,p+\dif p,  \overline \alpha;\Phi^{-\tau}z)\\
-\, &\rho(T+\dif T,p+\dif p,  \overline \alpha;z)\Big]   \dif z\ . 
\end{split}
\end{equation}
Thus, if the density $\rho$ tends (in weak sense)  to the 
final equilibrium density, with
 the internal parameters $\alpha$ adapted to the new external
ones, and if in addition it factors into an external part
 and an internal one (as should be assumed of  the initial density too),
then one should have 
$\Delta^\tau U \to \blangle H\brangle_{T+\dif T,p+\dif p}-\blangle
H\brangle_{T,p}$, i.e., 
$$
\blangle H\brangle_{T,p}
+\Delta^\tau U \to \blangle H\brangle_{T+\dif T,p+\dif p}\ ,
$$
where  $\blangle \cdot \brangle_{T,p}$
denotes mean value over the phase space of the system of interest, 
  with respect to the generalized  Gibbs
measure at values $p$, $T$ of the parameters.
 
So the time--dependent family $\Delta^\tau U$ considered here provides
an interpolation $\blangle H\brangle_{T,p} +\Delta^\tau U$ between the
initial equilibrium value $\blangle H\brangle_{T,p}$ and the final one $\blangle
H\brangle_{T+\dif T,p+\dif p}$ of the internal energy. However, it is
left to the dynamics to decide
whether, and at which time, will the final equilibrium value be
actually attained.  

\section{The second principle as a consequence of microscopic 
time--reversibility.  }\label{sec3}
The result proved in the previous section, according to which in
general   the
internal energy  of metaequilibrium thermodynamics
cannot be expressed as the mean of the Hamiltonian with
respect to the invariant measure, is the one to which we attach a
particular significance, as will be discussed in the appendix.
 In this section and in the next one, however, 
we continue the discussion of 
metaequilibrium thermodynamics. Here we  show how  the second principle
too is obtained. To this end we will show   that the coefficients
of the fundamental form $\delta Q-\delta W$ can be expressed  in terms
of suitable correlation functions, as in  FD theory. This fact is
obtained if  the
further assumptions  are made that i) the dynamics is time reversible,
and  thatii)  the  reservoirs are described by  Gibbs measures.
The latter condition can actually be weakened, as will be mentioned later.

So, let us recall the time--reversibility property.
This property amounts to requiring  that there exists a  mapping
$\mathcal P$ of the phase space onto itself
(inverting the sign of the velocities
of all particles that constitute the global system)  with the properties
$$
\mathcal {P}^2=\mathcal I\ ,\quad
(\mathcal{P}\Phi^\tau)^2 =\mathcal I\ ,   
$$
where $\mathcal I$ is the identity map. Obviously
the invariance will be required to hold also with respect to the
considered measure.
It is well known that the reversibility 
property  is satisfied by essentially all systems of interest, namely,
those with Hamiltonians  even in the momenta of the
particles. So in particular in our case we will have
\begin{equation*}
\begin{split}
& H_{\tot}(\mathcal{P}z)=H_{tot}(z)\ , \quad
 H(\mathcal{P}z)=H(z)\ ,\\
& H_{\ther}(\mathcal{P}z)=H_{\ther}(z)\ ,\quad \ldots \quad .
\end{split}
\end{equation*}
 
Concerning the family of measures for the initial data, we will assume
that the invariant measures have the form
\begin{equation}\label{sei}
\begin{split}
 \rho &(\beta,\beta_{\barra},\overline\alpha,z)\\
&=  \, c(\beta, \beta_{\barra}) \  e^{-\beta
   H_{\ther}}\  e^{-\beta_{\barra} H_{\barra}}\ 
\rho^{\syst} (\overline\alpha; z) \ ,
\end{split} 
\end{equation}
where $\rho^{\syst}$ refers to the system of interest,   $c(\beta, \beta_{\barra})$ 
is the familiar normalization factor involving the 
  partition functions of the reservoirs, while
$\beta$ and $\beta_{\barra}$ are Lagrange multipliers which determine
the temperature $T$ of the thermostat and the temperature  of the
barostat (and so implicitly its pressure  too). 

Our aim is now to find a suitable  expression for $W$ and thus for
$Q$. First of all the equilibrium pressure $p$ is defined as usual
(see \onlinecite{khinchin}) as the equilibrium mean value of the corresponding 
dynamical variable 
$ P(z)\equal  \frac {\partial H}{\partial  V}(z)=\frac 1 {A}\frac {\partial H}
{\partial l}(z)$. So the pressure $p$ is given by
$$
p= \blangle  P\brangle= \blangle \frac {\partial H}{\partial
   V} \brangle\ .
$$
Thus the work performed by the system of interest on the barostat should
be given by
\begin{equation}\label{lavoro}
W=p\blangle \Delta {V}\brangle^{\neqq}\ .
\end{equation}    
So  the heat $Q$ absorbed by the system of interest,  
$Q=\blangle \Delta H\brangle^{\neqq}+W$ (see (\ref{primo})), 
 will eventually  be expressed in terms of
dynamical variables of the system of interest, as
$$
Q=\blangle \Delta (H+p{V})\brangle^{\neqq}\ .
$$

We  thus turn to investigating the expression for the specific
heat. 
We even perform the more general computation
which gives the differential form $\delta Q$ in terms of $\dif \beta$
and $\dif \beta_{\barra}$.   Such a form  is
obtained from $Q$ through the same procedure used in the previous
section for the fundamental form $\delta Q-\delta W$, i.e., by
expanding the nonequilibrium density to first order in $\dif\beta$ and
$\dif\beta_{\barra}$. So one gets
\begin{equation}\label{formacal}
\delta Q= q_1(\beta,\beta_{\barra}) \,\dif \beta
+q_2(\beta,\beta_{\barra})\, 
\dif \beta_{\barra}\ , 
\end{equation}
where the  coefficients are given by  
\begin{equation}\label{coefcal}
\begin{split}
q_1&\equal  \int \Delta  (H+p{V})(z)\
\frac{\partial}{\partial \beta} \rho(\beta,\beta_{\barra},\overline\alpha; z) 
\dif z\ ,\\
q_2&\equal \int \Delta  (H+p{V})(z)\
\frac{\partial}{\partial \beta_{\barra}} \rho(\beta,\beta_{\barra},\overline\alpha;
z) \dif z\ .
 \end{split}
\end{equation}

One now has the following 
\par\noindent
Lemma: \emph{ Let the global Hamiltonian system be time--reversible, and let
  the invariant measure have a density  of the form (\ref{sei}). Then, 
for any dynamical variable $F$ having the property
\begin{equation}\label{simmetria}
F(\mathcal{P} z)= F( z)\ ,
\end{equation}
 one has}
\begin{equation}\label{derivate}
\begin{split}
\int \Delta F\ \frac{\partial}{\partial \beta}
\rho \dif z &=-\, \frac 12
\, \blangle \Delta  F\ \Delta (H+p{V})\brangle\\
\int \Delta F\ \frac{\partial}{\partial \beta_{\barra}}
\rho \dif z &= \frac p2\, 
\blangle \Delta  F \ \Delta{V}\brangle\ .
\end{split}
\end{equation}

The proof, for example for the first formula, is as follows. Due to the
exponential form of $\rho$ with respect to the reservoirs one has  
$$
I\equal\ \frac{\partial}{\partial \beta} \blangle \Delta  F \brangle
=-\, \int \Delta F (z) H_{\ther}(z)\rho(z) \dif z\ .
$$
Notice that, in performing the partial differentiation, one further term would
appear, which contains the derivative of the normalization factor
$c(\beta, \beta_{\barra})$. However, being  proportional 
to  $\blangle \Delta F\brangle$,
such a further term vanishes, in virtue of the invariance of the 
considered measure.
Introduce then the change of variables
$$
z=\mathcal{P}\Phi^\tau y
$$
which is canonical, so that $\dif z=\dif y$. Use  
(\ref{simmetria}), which implies
$$
\Delta F (z)=\, -\, \Delta F (y)\ ,
$$
and use also the invariance of the measure, $\rho
(z)=\rho(y)$. Thus, calling again by $z$ the dummy variable $y$ one
gets
$$
I =\int \Delta F (z) H_{\ther}(\mathcal{P}\Phi^\tau
z)\rho(z) \dif z\ .
$$
Finally, the result follows by  taking for $I$ the semisum of the 
original expression and of
the last one, using
$$
\Delta H_{\ther}=- \Delta (H+p{V})\ .
$$
The second formula is proved analogously, using
$\Delta H_{\barra}= p\, \Delta {V}$.

Thus the coefficients $q_1$, $q_2$ of the form $\dif Q$ turn out to be
expressed in terms of correlation functions with respect to the
invariant measure: 
\begin{equation}\label{coefcal2}
\begin{split}
q_1&= -\,\frac 12\,  \blangle\big[\Delta(H+p{V})\big]^2 \brangle  \\
q_2&=   \frac p2\, \blangle \Delta (H+p{V})\ \Delta 
{V}\brangle\ .
\end{split}
\end{equation} 
In particular, using  $\frac {\partial}{\partial T}=-(1/(k_BT^2) 
\frac {\partial}{\partial \beta}$  one finds that 
the specific heat $C_p$ at constant $\beta_{\barra}$ (i.e., at constant
$p$),  is manifestly positive,
\begin{equation}\label{calspec}
C_p=\frac 1 {2k_BT^2}\blangle\left[\Delta(H+p{V})\right]^2
\brangle\, \ge 0\ .
\end{equation}
Thus  the second principle in the
Clausius form (heat spontaneously flows from hot to cold bodies) is
proven. By the way, one is naturally led to guess that the second
principle may not hold if microscopic time--reversibility were not satisfied. 

The positiveness of the specific heat was proven above, making use of
the assumption that the marginal measure for the environment be of
Gibbs type. However, if one looks  at the proof one realizes that the
result also holds if any  assumption is made which guarantees that
the partial derivatives of the density $\rho$ have negative definite
sign.  For example, a sufficient condition is that, in
formula (\ref{sei}) for the density,  instead of the exponential $\exp
( -\beta H)$ there appears a factor $f(\beta H)$, with any function $f$
having the property   $f'\le 0$.

Having thus proven  positiveness of the
specific heat, even in such a more general frame,
the existence of an integrating denominator for the
form $\delta Q$  then follows from    arguments of 
Carath\'eodory type. A connection with the temperature of the
thermostat will be provided in the next section. 

Finally, the fact that the thermodynamic quantities may attain the
corresponding equilibrium values if a suitable relaxation takes place, 
is exhibited by expressing the thermodynamic coefficients  in terms of
time correlation functions. For example, in the case of the specific
heat  $C_p$ given by (\ref{calspec}),  by expanding the square
$$
\left[ \Delta (H +pV)(z)\right]^2=\left[(H+pV) (\Phi^\tau z)-(H+pV)(z)\right]^2
$$ 
and adding and subtracting the term $\blangle H+pV\brangle^2$
one immediately gets
\begin{equation}\label{dieci}
C_p=\frac{1}{k_BT^2}\left[\sigma^2_{H+pV} \, -\, 
\mbox{\calligra C}_{H+pV}(\tau) \right]\ ,
\end{equation}
where $\sigma^2_F$ denotes the variance of $F$ and 
$ \mbox{\calligra C}_F(\tau)\equal \blangle F(\Phi^\tau
z)F(z)\brangle - \blangle
F\brangle^2$  its time--autocorrelation function.

\section{The temperature of the thermostat as an integrating
  denominator for $\delta Q$}\label{sec4}
Thus, having proven the second principle, we know that an integrating
denominator for the differential form $\delta Q$ exists. One may ask
however whether such  an integrating denominator coincides with the
temperature  $1/\beta$ of the thermostat. We will show now that this is actually
the case, at least if the observation time $\tau$ is  sufficiently
large to guarantee that a suitable dynamical decorrelation did occur.

So, let us  consider the form \begin{equation}\label{forcal/t}
\beta {\delta Q}=  \beta {q_1}  \,\dif \beta +\beta {q_2} \, \dif\beta_{\barra}\ , 
\end{equation}
where the coefficients $q_i$ are defined by (\ref{coefcal}), and let
us investigate whether it is closed. This requires to  calculate 
$\frac{ \partial(\beta q_1)}{\partial \beta_{\barra}}-\frac {\partial (\beta
    q_2)}{\partial\beta}$. As in the case of the internal energy,
again one finds that the two terms involving the second derivatives of
$\rho$ cancel, so that one remains with
\begin{equation}
\begin{split}
&\frac{ \partial(\beta q_1)}{\partial \beta_{\barra}}-\frac {\partial (\beta
    q_2)}{\partial\beta}\\
&= \beta \frac{\partial p}{\partial\beta_{\barra}}\int \Delta  {V}(z)\ 
\frac{\partial}{\partial \beta}\, \rho\, \dif z\\
& -\int \Delta (H+p  {V})(z)\ 
\frac{\partial}{\partial \beta_{\barra}}\, \rho\, \dif z \ .
\end{split}
\end{equation}
So, using the Lemma one gets
\begin{equation}
\begin{split}
& \Big[\frac{ \partial(\beta q_1)}{\partial \beta_{\barra}}-
\frac {\partial (\beta    q_2)}{\partial\beta}\Big]\\
&=-\, \frac 12\,   \blangle  \Delta  {V}\ 
\Delta (H+pV)\brangle\ \left(\beta \frac{\partial
  p}{\partial\beta_{\barra}} +  p \right) \ .
\end{split}
\end{equation}
Thus the form is closed if one has
\begin{equation}\label{chiusura}
\frac{\partial p}{\partial\beta_{\barra}}= -\, \frac p {\beta} \ .
\end{equation}

To investigate this point requires to find the expression of 
$\frac{\partial p}{\partial\beta_{\barra}}$, and this is obtained by
finding the expression of $\dif p$ as a differential form with respect
to $\dif \beta$, $\dif \beta_{\barra}$. In analogy with what previously
done for other differential forms, we just define $\dif p$ as the
first order expansion of $\blangle \Delta  P\brangle^{\neqq}$
with respect to $\dif \beta$, $\dif \beta_{\barra}$. This gives, using
again the Lemma,
\begin{equation}\label{dipi}
\dif p=-\frac 12\blangle \Delta P \Delta (H+pV)\brangle \, \dif \beta
+ \frac p2\blangle \Delta  P \Delta V\brangle\, \dif \beta_{\barra}
\end{equation}
from which one gets
\begin{equation}\label{chiusura2}
\frac{\partial p}{\partial\beta_{\barra}}= 
\frac p2\blangle \Delta P \Delta V\brangle \ .
\end{equation}
On the other hand, if a suitable decorrelation occurs, one has as usual
\begin{equation}\label{chiusura3}
\frac 12 \,\blangle \Delta P \Delta V\brangle\to 
 \blangle  \left(P -\blangle P\brangle\right)
\left( V -\langle V\brangle\right)   \brangle\ .
\end{equation}
Finally,  the r.h.s. of (\ref{chiusura3}) is just equal to $ -\, 1/\beta$, as
one finds in the Landau--Lifshitz textbook \cite{landau}.  

Thus after a sufficiently long time the temperature of the thermostat
is an integrating denominator of the form $\delta Q$. It is presumable
however that such a relaxation time might in general be rather short,
especially in comparison with the relaxation times  related to 
the specific heats at low temperatures.

\section{Conclusions}\label{sec6}
In this paper we have shown how it is possible to formulate
 statistical thermodynamics for a Hamiltonian system, in terms of the  
processes  induced by a change 
of the thermodynamic parameters.  
A characteristic feature of our approach is that it  allows one 
to deal with metaequilibrium (or partial equilibrium) situations in
which the time--correlations did not decay to zero, in addition to
the equilibrium situations  described in the famililar way through
equilibrium ensembles. 
An approach somehow similar to ours  was recently proposed by a 
C. Bernardin and S. Olla
\cite{olla}. The main
difference is that  we frame the problem in a time--independent
Hamiltonian setting in which the processes are induced by
nonequilibrium  measures for the initial data, whereas in the quoted
paper the processes are induced by imposing suitable time
dependencies on  the parameters of the global system.

A relevant  feature of the present  approach
 is that it involves in a substantial way, in addition
to the system of interest, also the measurement
apparatuses. Consequently, one has the problem of understanding  
whether a  physical, actually measured,
property  belongs to the system of interest 
or rather  depends on the  measuring procedure too. 
Standard equilibrium statistical
thermodynamics, being  formulated in terms of the Hamiltonian of the system
of interest only, avoids in principle any reference to the method of
measurement, and perhaps one would be tempted to say that this is 
indeed correct. In general, however, things are not so. And actually,
in the work \onlinecite{birge}, of which it was said that it gave rise to a
new epoch of thermodynamics,
one even finds the sentence: \emph{``How does one interpret a quantity
  such as the specific heat in a nonequilibrium situation?
Clearly, the dynamics of how one performs the measurement enters into
the measured value''}. This circumstance finds a
counterpart in the present approach. Indeed the thermodynamic quantities we
are considering, typically the specific heat of the system of
interest, are in general represented by  expressions which, through the
intermediate of the  measure, involve not only the system of interest,
but also the sources. So the measured quantities  in general depend on the
measurement procedure, and the dependence disappears only when  a suitable 
relaxation has occurred, if  indeed it takes place.

However, the main interest of  the proposal made here  is that it 
may  provide an
 answer to a question
of principle that remains open  within the equilibrium approach. The
problem is to understand how many degrees of freedom  should be
attributed to  a  microscopic model.  Why can one neglect 
the rotational degrees of freedom in a system of hard, completely 
smooth, spheres?  Conversely, why does a supercooled liquid  apparently
have a number of degrees of freedom larger than the corresponding
crystal?  

Problems of this type are usually met in  chemistry or 
in glasses (see for example \onlinecite{parisi}). 
In fact, however, they touch   on  questions 
of principle
that afflicted very much  Boltzmann (see \onlinecite{nature} and the
quotations reported in \onlinecite{noi}) and Gibbs \cite{gibbs}, and were
later taken up
by authors such as  Rayleigh and  Jeans (see the first Solvay
Conference \cite{solvay}, particularly the letter sent by Rayleigh) and
Nernst \cite{nernst}, in connection with foundational problems of quantum
mechanics. 

In an appendix it is discussed  how the
present results may be of interest for the
Fermi--Pasta--Ulam problem, which also presents metaequilibrium
situations (as apparently first suggested  by Parisi and
collaborators\cite{parisi2}),  and  thus constitutes a modern arena
for dealing with foundational problems of the type just mentioned.

\appendix

\section*{Appendix}
\textbf{Possible Relevance of Metaequilibrium Thermodynamics 
for the Fermi--Pasta--Ulam Problem.}
In this paper we have shown how a statistical thermodynamics different from the 
equilibrium one can be formulated in situations in which the system does not
present a full dynamical relaxation, i.e.,  in situations of  partial
equilibrium, as  Landau and Lifshitz would say. 
This can occur even if the system
was originally in an equilibrium state, described by an invariant  
Gibbs measure. In particular, if the dynamics is such
that the decorrelation of $H+pV$ is negligible for long times when 
$T\to 0$, i.e., if one has $\mbox{\calligra C}_{H+pV}(\tau)
\simeq \sigma^2_{H+pV}$ for long times when  $T\to 0$,
then one has $C_p\simeq 0$, and so the  system behaves 
in agreement with the third principle. 

Situations of this   type were  conceived by Boltzmann since his
celebrated paper  published in the journal Nature\cite{nature} in the year 1895
(see also the
quotations reported in  \onlinecite{noi}) and were  amply discussed
in the year 1916 by Walther Nernst\cite{nernst},
who came to the point of  conceiving
 that equipartition of energy might be compatible with
Planck's law. According to Nernst, the key point consists in
distinguishing between the
mechanical  energy (which corresponds to equipartition if the initial
data are Gibbs distributed), and the energy that can   be dynamically
exchanged in a process. Only the exchangeable energy
should be identified with the thermodynamic internal energy $U$,
while the complementary part should play the role of a kind of
``zero--point energy''.  

Foundational questions of this type were reconsidered later in the
context of the FPU problem.   
As previously mentioned,  the idea that in the FPU problem one meets
with metaequilibrium situations analogous to those of glasses was
first  advanced by a group of people about Parisi in the
year 1982 \cite{parisi2} (see also \onlinecite{parisi3} and the review 
\onlinecite{fpurev}).

In the FPU problem one has a classical approach, in which initial data
are considered with only some of the normal modes excited (typically,
the low frequency ones), and the result found is that, for low enough
specific energies, up to extremely long
times energy remains confined  among a certain group of low frequency modes. Already in the
year 1972, in a paper of Cercignani et al. \cite{cerci} by the title 
\emph{``Zero--point energy in   classical non--linear mechanics''} 
it was however proposed
 that even in the FPU problem  one might meet with
situations of the type  conceived  by Nernst.

Numerical studies on the variant of the FPU problem with Gibbs
distributed initial data came later, with results that were variously
interpreted (see \onlinecite{fiorentini}, \onlinecite{tenenbaum}). 
Finally a case was considered, which involves a situation very
similar to the one discussed here \cite{cg}, where one measures 
the specific heat of  an FPU system (with an initial Gibbs
distribution),  put in contact with a heat reservoir. 
The result found is that at high 
enough energies a
complete decorrelation occurs, and the standard equilibrium value
is measured. Instead, at low enough energies situations of an apparent
metaequilibrium (or partial equilibrium) are met, and smaller
values of the specific heat are  measured. Situations of metastability with
hysteresis effects had been previously reported \cite{cghysteresis}. 
Finally, the time
autocorrelation function of the mode energies, or of the energies of
packets  of nearby frequencies, were numerically  studied in the paper 
\onlinecite{corre}.

So much for what concerns numerical results.
In the meantime a consistent progress was achieved at an analytical
level, because it was proven that a weak version of perturbation
theory can  be formulated in the thermodynamic limit \cite{andrea}. 
At variance with
the classical approach, one renounces to control all  initial data, and
only looks at quantities such as the time--autocorrelation functions of
suitable dynamical variables, estimated with respect to the Gibbs
measure. This allows one to obtain results which hold
in the thermodynamic limit of
infinitely many particles, with a finite temperature. 
It was found  that such autocorrelation functions do not
decay to zero up to rather long times:  
a Nekhoroshev--like  exponential
stability was  found for a $\Phi^4$--type model \cite{aa}, while one step of
perturbation theory was performed in the very FPU model \cite{baa}.
Such results appear to disprove the common expectation (see for
example \onlinecite{olla}) that situations of partial equilibrium 
as those discussed here should not occur for infinite systems.

Concerning the connections with the   foundations of quantum
mechanics,  we now
add a comment on the  work of Poincar\'e of the year 1912 on
the necessity of quantization \cite{poincare} (see also the work
\onlinecite{ehrenfest} of Ehrenfest). 
The problem was posed in the following way. Assume  that the internal
energy $U$ of the system of interest (the blackbody) be expressed as
 the mean value of the Hamiltonian $H$ with respect to a
suitable measure,
\begin{equation}\label{poincare}
U=\int H(z)\,  \sigma(z)\, \dif z\ .
\end{equation}
Then, it is asked which qualitative properties should the 
measure $\sigma$ have in order 
that the  function $U$ given by (\ref{poincare}) agree with  the
phenomenological internal energy. The answer of  
Poincar\'e  was that, under quite
general  assumptions, quantization is necessary, i.e., the measure
has to be singular.
 
Our comment is that, if one takes the metaequilibrium perspective,   
the claim that  $U$ 
 should be expressed as an equilibrium  mean
 value of the energy $H$  (rather than a nonequilibrium mean value of
 $\Delta^\tau H$), is unfounded.  The result of
 Poincar\'e shows  that,   if one does so,
 i.e., if  one insists in   representing
 $U$  as an equilibrium
 mean value of $H$, then everything goes as if energy were
 quantized. Or, in Einstein's words at his
 contribution to the first Solvay Conference
 (see \onlinecite{einstein}, section 3), 
 \emph{``The statistical properties of the phenomenon are the same as
   if energy were transferred only by integer quanta of size $h\nu$''}. 

Curiously enough, an opinion similar to the latter one of Einstein
  was expressed by Poincar\'e
too,  in a paper of a less technical character \cite{poincare2}
 written shortly after the  previously mentioned one, where he says:
\emph{``Will discontinuity reign over the physical universe and will its
triumph be definitive? Or rather will it be recognized that such a
discontinuity is only an appearance and that it dissimulates a series
of continuous processes? The first person that saw a collision
believed to be observing a discontinuous phenomenon, although we know
today that the person was actually seeing the effect of very rapid
changes of velocity, yet continuous ones.''}  And the conclusion was:
\emph{ ``To try to express  today an opinion about these problems would
  mean to be wasting one's  ink.''}

Now, our admiration for Poincar\'e is unlimited, but our personal
feeling, or rather hope,  is that perhaps on this point Einstein was
seeing farther than him.  It appears that  the main
intuition of Einstein about the role of fluctuations was correct,
although \emph{``rather ironically''}  
(to use the words of  \onlinecite{nielsen})
 he couldn't have   a full appreciation of their dynamical
features, which was obtained with FD theory only  half a
century later. It is just the stabilization of the
time--autocorrelation functions to well definite nonvanishing values
that allows for a thermodynamics different from the equilibrium one
to hold. And this,  not only for chemical systems or for  glassy--like  ones, but perhaps also
for systems of FPU type, 
which are of interest for the foundations of physics. By the way, if
one looks at the objections raised by Poincar\'e againsts the
``partial or incomplete equilibrium'' point of view of Rayleigh and
Jeans, one realizes that what he could not conceive  is just that such
partial equilibrium situations may actually occur in nature, and that
they  may  be dealt with  through the methods of  phenolemological 
thermodynamics.

\end{document}